\documentclass{webofc}

\usepackage[varg]{txfonts}   
\usepackage{hyperref}
\usepackage{url}
\hypersetup{colorlinks=true,citecolor=blue,urlcolor=blue,linkcolor=blue}
\begin{document}
\title{MC-EKRT: Monte Carlo event generator with saturated minijet
production for initializing 3+1 D fluid dynamics in high energy
nuclear collisions}

\author{\firstname{Harri} \lastname{Niemi}\inst{1,2} \and
        \firstname{Jussi} \lastname{Auvinen}\inst{1,2} \and
        \firstname{Kari J.} \lastname{Eskola}\inst{1,2} \and
        \firstname{Henry} \lastname{Hirvonen}\inst{3} \and
        \firstname{Yuuka} \lastname{Kanakubo}\inst{4,5} \and
        \firstname{Mikko} \lastname{Kuha}\inst{1,2}
        }

\institute{University of Jyv\"askyl\"a, Department of Physics, P.O.B. 35, FI-40014 University of Jyv\"askyl\"a, Finland \and
          Helsinki Institute of Physics, P.O.B. 64, FI-00014 University of Helsinki, Finland \and
          Department of Mathematics, Vanderbilt University, Nashville TN 37240, USA \and
          RIKEN Center for Interdisciplinary Theoretical and Mathematical Sciences (iTHEMS), RIKEN, Wako 351-0198, Japan  \and
          Nuclear Science Division, Lawrence Berkeley National Laboratory, Berkeley, CA, 94720, USA
          }

\abstract{We present a novel saturation and leading order collinear factorization based Monte-Carlo implementation of the EKRT model for computing QCD matter initial states in high-energy nuclear collisions. As new features the MC implementation gives a full 3-dimensional initial state event-by-event, introduces new event-by-event fluctuating spatially dependent nuclear parton distribution functions, includes dynamical fluctuations in minijet production and saturation, and accounts for the energy-momentum and valence-quark number conservation.
}
\maketitle
\vspace{-1cm}
\section{Introduction}

Monte-Carlo EKRT model (MC-EKRT) is a novel framework to compute initial conditions for hydrodynamical evolution of QCD matter in nuclear collisions \cite{Kuha:2024kmq, Hirvonen:2024zne}. It is based on leading-order (LO) collinearly factorized perturbative QCD (pQCD) and a saturation conjecture which controls the production of low-$p_T$ partons. As such, it is an extension of the EKRT model (EbyE-EKRT) that has been very successful in describing a wealth of low-$p_T$ observables at midrapidity measured in Pb+Pb and Xe+Xe collisions at the LHC, and 200 GeV Au+Au collisions at RHIC \cite{Niemi:2015qia, Niemi:2015voa, Eskola:2017bup, Hirvonen:2022xfv}. As major new features, the MC-EKRT model has now dynamical fluctuations in the particle production, i.e.\ fluctuations that are not coming from the geometrical fluctuations in the initial positions of the nucleons inside the nuclei, and it gives full 3-dimensional rapidity dependent initial conditions.

\section{MC-EKRT}

An essential quantity in MC-EKRT is the (mini-)jet cross section $\sigma_{\rm jet}$ computed from the LO collinearly factorized pQCD. In computing $\sigma_{\rm jet}$ we employ new spatially dependent event-by-event (EbyE) fluctuating nuclear parton distribution functions (nPDF), where the spatial dependence and EbyE fluctuations come from the fluctuations in the nucleon configuration in each nucleus. The main mechanism that controls the low-$p_T$ parton production, and thus the total produced energy and eventually the energy density profiles, is saturation. Here we employ a Monte-Carlo implementation of the EKRT saturation conjecture \cite{Eskola:1999fc}. The MC-EKRT simulation gives then as an output a partonic final state with full 3D kinematics for all the partons. This state can then be converted to an initial energy density profile for a (3+1)D viscous hydrodynamical evolution.

The main idea behind the EKRT saturation conjecture is that saturation occurs when initial gluon density becomes so large that  $3\rightarrow2$, $4\rightarrow2$, etc.\ partonic processes become as important as $2\rightarrow2$ processes \cite{Paatelainen:2012at}. This conjecture leads to a scaling law that can be in a central ($b=0$) collision cast in a form \cite{Eskola:1999fc},
\begin{equation}
N_{AA}^{2\rightarrow 2}(0)\frac{\pi}{p_0^2} \sim \pi R_A^2,
\label{E:EKRTsat}
\end{equation}
where $N_{AA}$ is the number of $2\rightarrow2$ processes, $R_A$ is the nuclear radius, and $p_0$ is the lower $p_T$ cut. Here
$\pi/p_0^2$ can be interpreted as a transverse formation-area for a produced dijet \cite{Eskola:1999fc}. Thus, the minijet production saturates when the minijet production processes fill the whole available transverse plane. In MC-EKRT this condition is implemented by requiring that partonic processes with a transverse area $\pi/p_T^2$ cannot overlap each other in the transverse plane. In practice each candidate minijet process is compared to already accepted higher-$p_T$ processes, and it is rejected if for the transverse distances $\bar{s}$
 \begin{equation}
        |\bar{s}-\bar{s}^{\text{cand}}| < \frac{1}{\kappa_{\mathrm{sat}}}\left(\frac{1}{p_T}+\frac{1}{p_T^{\text{cand}}}\right),
\label{E:saturation}
\end{equation}
where $\kappa_{\rm sat}$ is a free parameter that controls the strength of the saturation.

The nuclear effects in the PDFs are implemented as modification factors $r_i^{a/A}$, so that nPDF is given by
\begin{equation}
f_i^{a/A}(\{\bar{s}_a\},x,Q^2)  = f_i^{a}(x,Q^2) r_i^{a/A}(\{\bar{s}_a\},x,Q^2),
\end{equation}
where $f_i^{a}(x,Q^2)$ are the free proton PDFs. The $r_i^{a/A}$ factors depend on the nucleon configuration through the nuclear overlap function $\hat{T}_{A}^a$ (averaged over nucleon transverse area for each nucleon separately), and they are constructed in such a way that when averaged over a large sample of nucleon configurations $\{\bar{s}_a\}$ they give back the EPS09 nPDFs \cite{Eskola:2009uj}.

The main steps in the MC-EKRT simulation of nuclear collisions are the following:

\noindent\textbf{1.} First the impact parameter and positions of the nucleons in each nucleus are sampled, and a triggering condition for an inelastic nuclear collision is checked.

\noindent\textbf{2.} If the collision is triggered the number of candidate dijets is sampled from a Poissonian probability distribution, with an average number of dijets in each nucleon-nucleon pair being $T_{nn}(b) \sigma_{\rm jet}$, where $T_{nn}(b)$ is the nucleon-nucleon overlap function, $b$ is the $nn$ impact parameter, and $\sigma_{\rm jet}$ is computed from LO collinear factorization including collision-energy dependent $K$-factors as free parameters that account for the NLO contributions.

\noindent\textbf{3.}
The parton flavors and their momenta are sampled from the LO pQCD cross sections, and the parton transverse positions are sampled from the nucleon-nucleon overlap function.

\noindent\textbf{4.}
The excess dijet processes are then filtered in the order of decreasing $p_T$ using the saturation criterion. The remaining
dijets are then further filtered, again in the order of decreasing $p_T$ using the per-nucleon momentum conservation and valence-quark number conservation.

\noindent\textbf{5.}
The transverse energy $E_T$ of the remaining minijets is then converted into a local energy density profile using a Gaussian smearing.

\noindent\textbf{6.}
The subsequent evolution is computed using (3+1)D relativistic hydrodynamics \cite{Molnar:2014zha}, and the final hadron spectra are then computed using the Cooper-Frye freeze-out procedure.

\begin{figure*}
\centering
\includegraphics[width=10cm,clip]{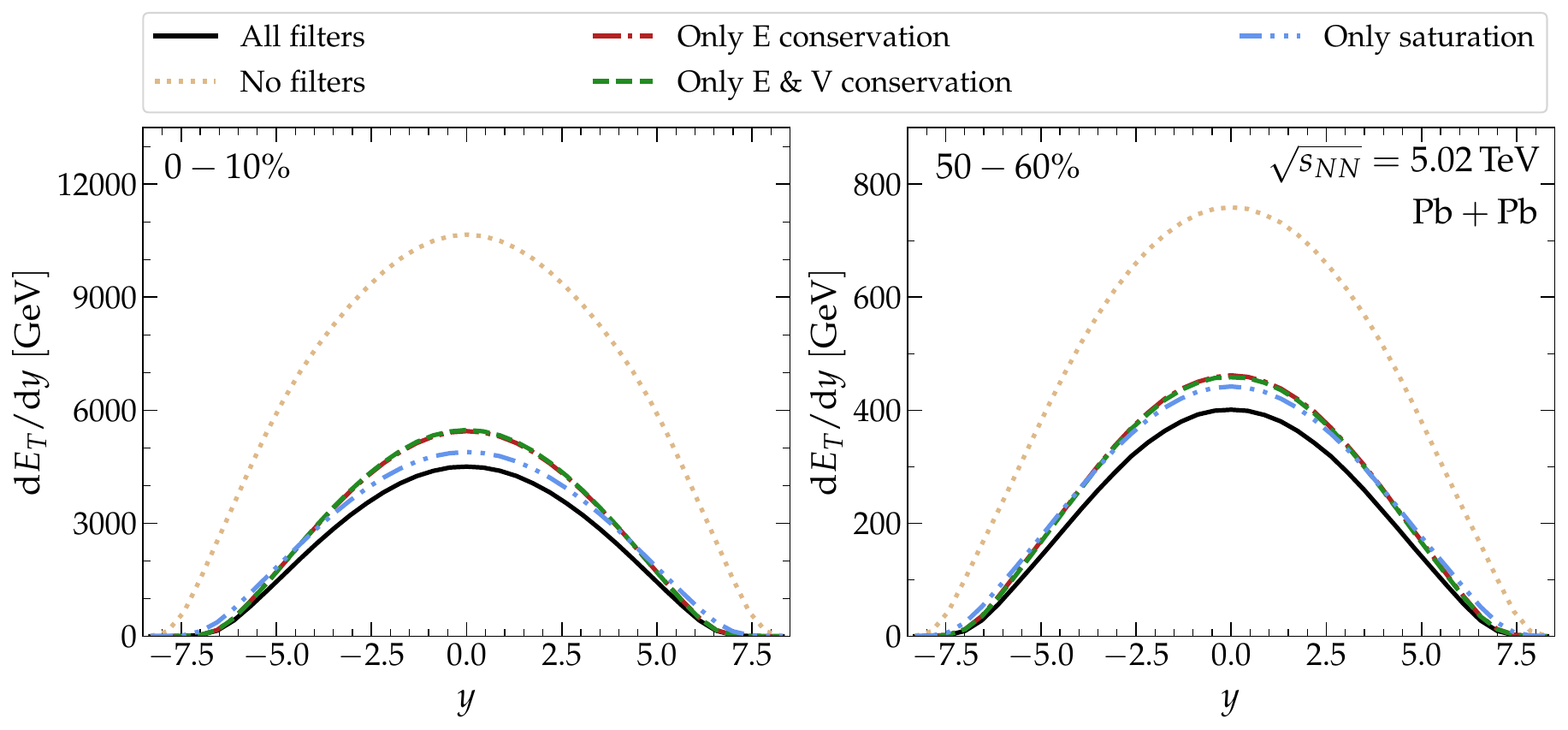}
\vspace*{-0.1cm}
\caption{The impact of the EKRT saturation, energy-conservation and valence-quark number conservation filters on the rapidity dependence of minijet transverse energy in $\sqrt{s_{NN}}=5.02\,$TeV central (left panel) and peripheral Pb+Pb collisions (right panel). Figure is from Ref.~\cite{Kuha:2024kmq}.}
\label{fig:filter}       
\vspace*{-0.92cm}
\end{figure*}

The effect of the filters is shown in Fig.~\ref{fig:filter}. As can be seen from the figure the saturation filter is the dominant effect that suppresses the transverse energy of the candidate dijets. After saturation the energy-momentum conservation has a relatively small effect, and the valence-quark conservation has even smaller impact on the transverse energy.

We show below the first preliminary results for hadron multiplicity and elliptic flow from the MC-EKRT model. In order to speed up the computation of full (3+1)D fluid evolution, we first average a large number of initial entropy density profiles for each centrality class, where the centrality is determined from the total minijet transverse energy, and compute the hydrodynamical evolution only for these averaged initial states.

\section{Results}

\begin{figure*}
\centering
\includegraphics[width=6.0cm,clip]{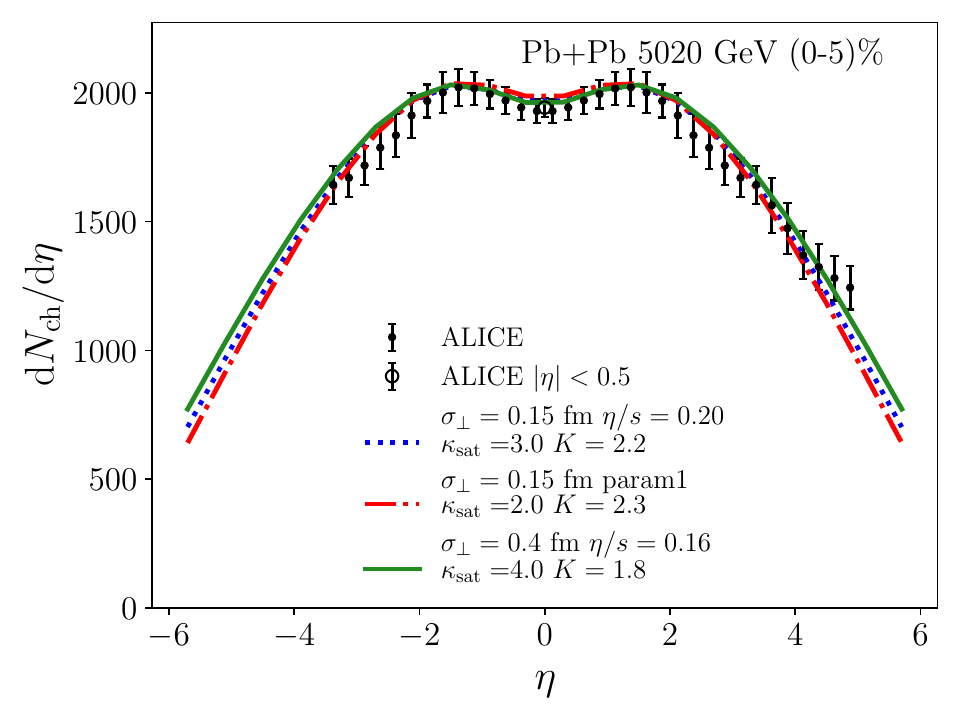}
\includegraphics[width=6.0cm,clip]{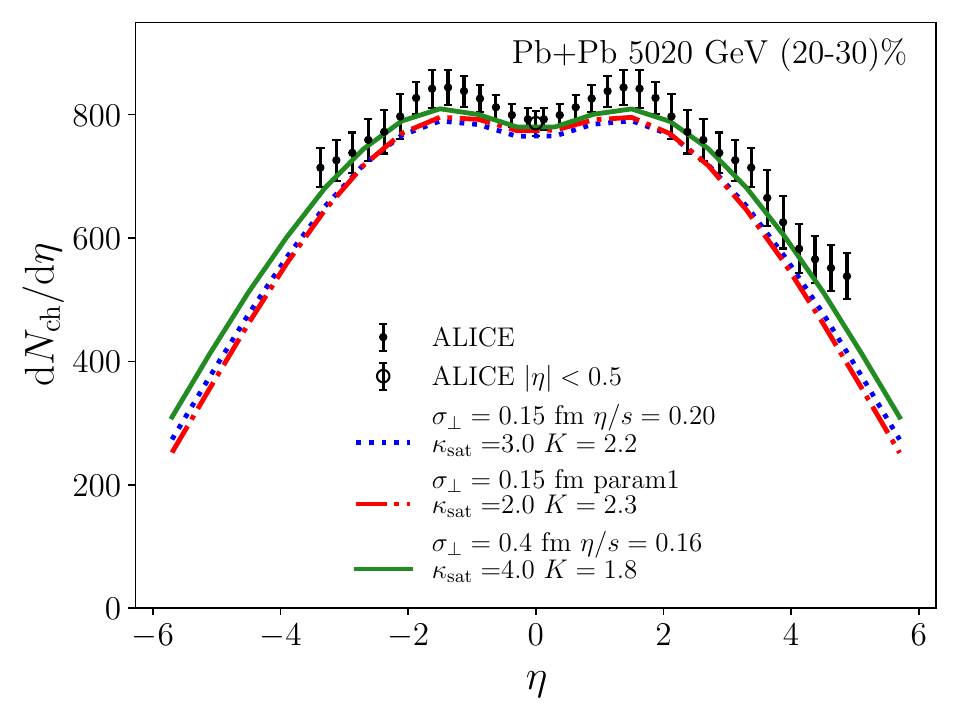}
\includegraphics[width=6.0cm,clip]{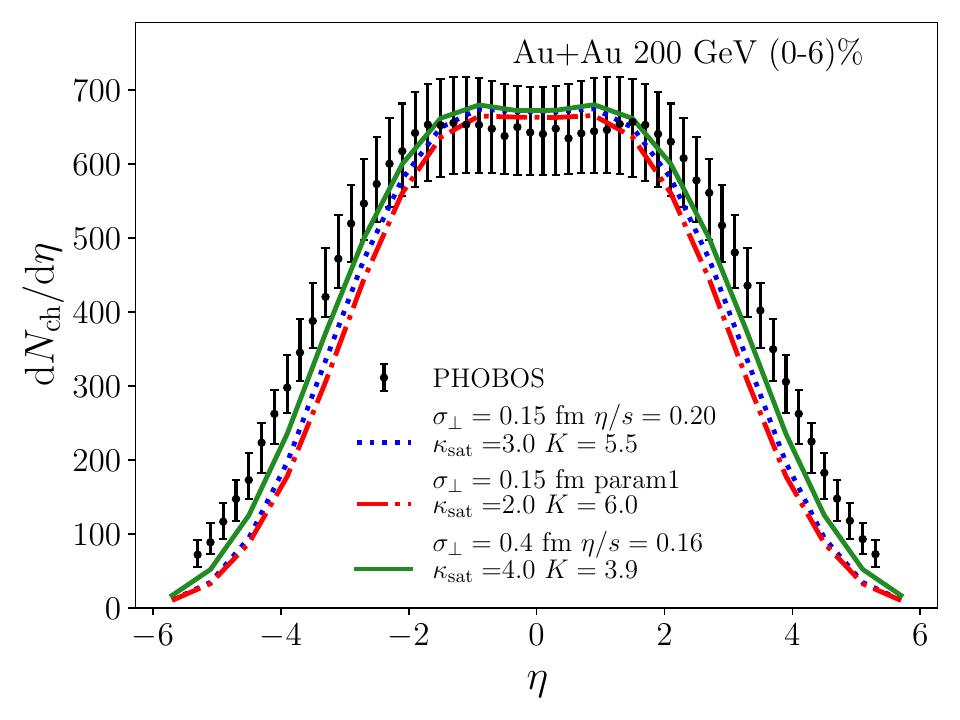}
\includegraphics[width=6.0cm,clip]{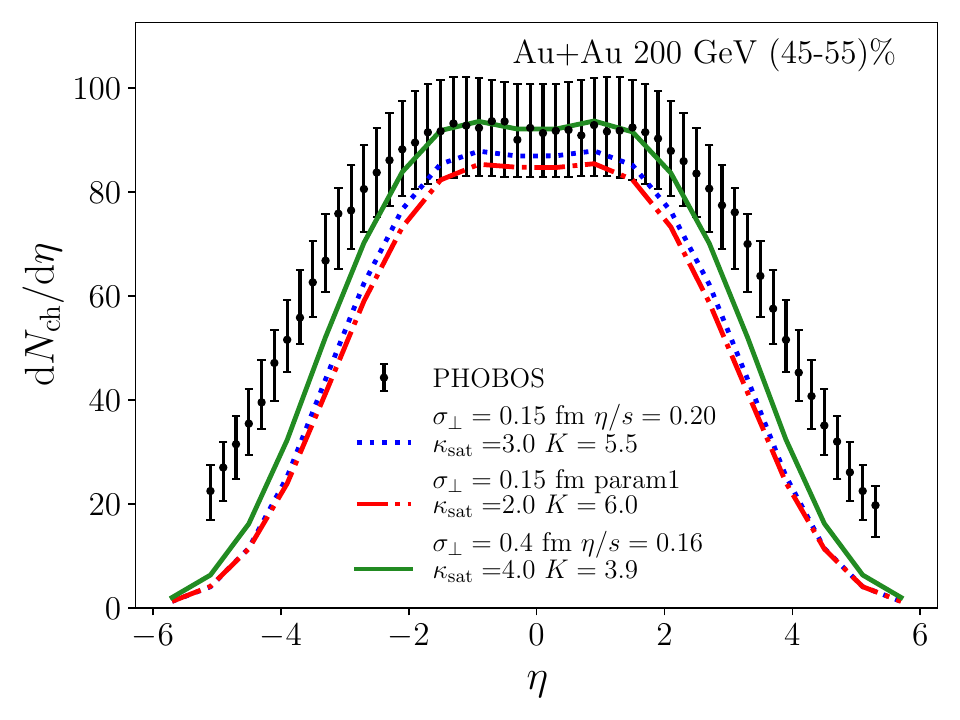}
\includegraphics[width=6.0cm,clip]{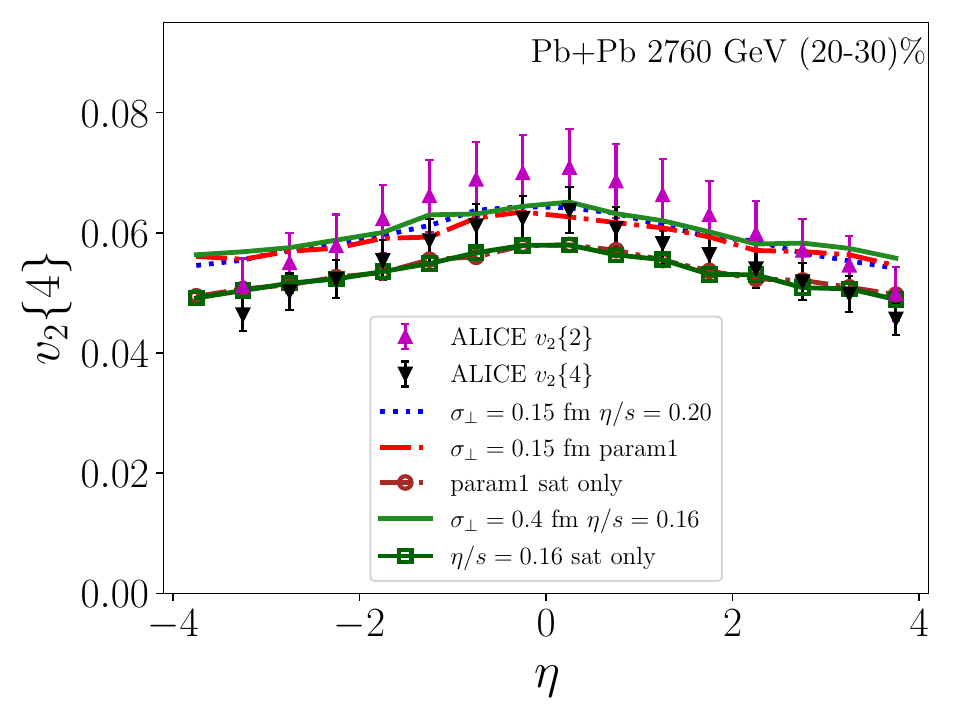}
\includegraphics[width=6.0cm,clip]{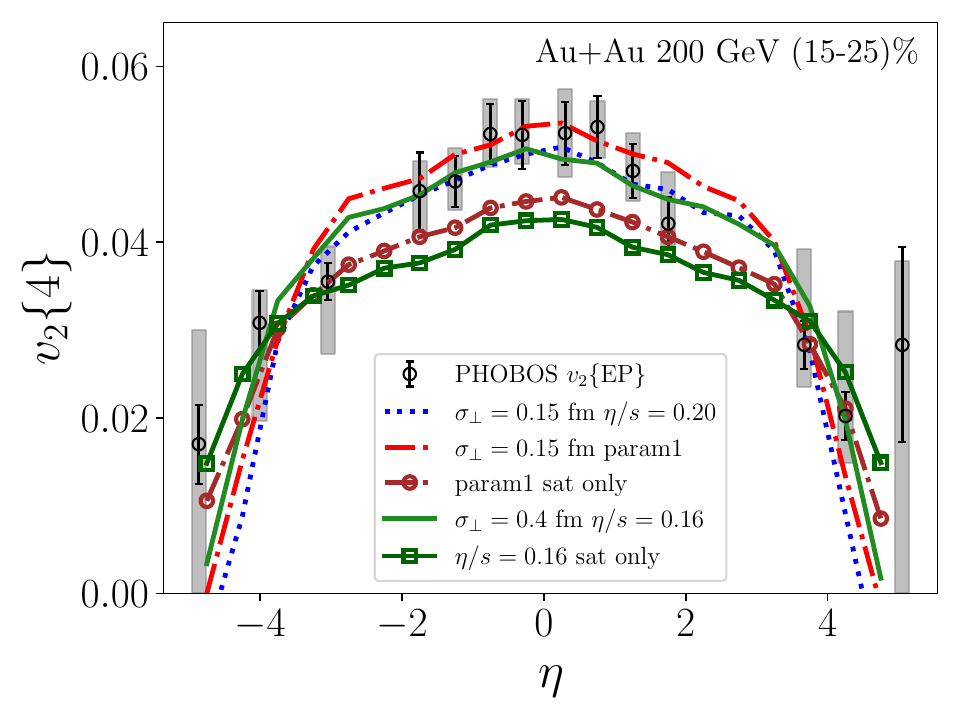}
\vspace*{-0.4cm}
\caption{Charged particle multiplicity $\mathrm{d}N_{\mathrm{ch}}/\mathrm{d}\eta$ as a function of pseudorapidity in Pb+Pb collisions at $\sqrt{s_{NN}}=5.02$ TeV, compared with ALICE data~\cite{ALICE:2016fbt,ALICE:2015juo} (top panels),
in Au+Au collisions at $\sqrt{s_{NN}}=200$ GeV, compared with PHOBOS data~\cite{Back:2002wb} (middle panels).
The bottom panels show the charged particle $v_2\{4\}$ as a function of pseudorapidity in $\sqrt{s_{NN}}=2.76$ Pb+Pb collisions, compared with ALICE data~\cite{ALICE:2016tlx} (left), and in $\sqrt{s_{NN}}=200$ GeV Au+Au collisions compared with PHOBOS event plane $v_2$ data~\cite{PHOBOS:2004vcu}. Figures are from Ref.~\cite{Kuha:2024kmq}.}
\vspace*{-0.5cm}
\label{fig:results}       
\end{figure*}

In Fig.~\ref{fig:results} we show the pseudorapidity distribution of hadrons in 5.023 TeV Pb+Pb and 200 GeV Au+Au collisions, the data are from the ALICE \cite{ALICE:2016fbt,ALICE:2015juo} and PHOBOS Collaboration \cite{Back:2002wb}, respectively. The top two left panels show the central collisions and right panels semi-peripheral collisions. The lowest two panels show the pseudorapidity dependence of elliptic flow in 2.76 TeV Pb+Pb collisions, and in 200 GeV Au+Au collisions, where the data are from the ALICE \cite{ALICE:2016tlx} and PHOBOS Collaborations \cite{PHOBOS:2004vcu}, respectively.

The overall agreement with both the centrality and pseudorapidity dependence of the measurements is very good, only at the larger rapidities in peripheral collisions at RHIC do we start to see larger deviations. This suggests that the same saturation mechanism that has successfully explained the mid-rapidity observables, works well also at larger rapidities.

{\it Acknowledgments.} This research was funded as a part of the Center of Excellence in Quark Matter of the Academy of Finland (Projects No. 346325 and 364192). This research is part of the European Research Council Project No. ERC-2018-ADG-835105 YoctoLHC. We acknowledge the computation resources from the Finnish IT Center for Science (CSC), project jyy2580, and from the Finnish Computing Competence Infrastructure (FCCI), persistent identifier urn:nbn:fi:research-infras-2016072533.

\end{document}